# Evidence of enhanced Zn-diffusion observed during the growth of Inverted Metamorphic Solar Cells

Manuel Hinojosa, Iván García, Ignacio Rey-Stolle and Carlos Algora

Instituto de Energía Solar, ETSI de Telecomunicación,
Universidad Politécnica de Madrid, 28040 Madrid, Spain

*Abstract* — Zinc-diffusion can induce multiple failures in the electrical performance of a multijunction solar cell. In this work, we show an important Zn-diffusion from the AlGaInP back-surface-field layer to the emitter of the GaInP top cell of an inverted multijunction solar cell. Through the analysis of different doping profiles, we provide strong evidence that the diffusion mechanism is (1) triggered by the growth of the tunnel junction cathode and (2) involves point defects. We analyze the implications of Zn-diffusion on the bandgap, the rear-passivation and the minority carrier quality of the GaInP solar subcell by relating the electrical performance of different samples to its corresponding doping profile.

## I. INTRODUCTION

Inverted Metamorphic Multijunction (IMM) Solar Cells have become an attractive pathway to obtain high-efficiency solar cells, as they allow to combine a wide range of materials with different bandgaps and lattice constants in a single epitaxial growth [1]. However, the growth of semiconductor structures in the inverted direction (i.e. n-type layers first) presents remarkable differences with respect to the upright ones (i.e. p-type layers first). For example, in an inverted growth, the front layers of the solar cell suffer the thermal annealing derived from the epitaxial growth of the remainder of the structure, which typically takes some hours at temperatures from 500 to 700ºC, meaning that diffusion of impurities and point defects in the structure occurs under significant different conditions [2].

During the development of a triple-junction IMM GaInP/GaAs/GaInAs solar cell in our MOVPE reactor, we observed a heavy alteration in the doping concentration of the GaInP subcell when it was integrated as a top subcell (TC) into a 2 and 3-junction IMM structure. The origin of such variation was a strong Zn-diffusion away from the p-type AlGaInP back-surface-field (BSF) towards the rest of structure during the growth of the rest of the structure, resulting in a reduction of the p-type doping level in the BSF and a partially compensated n-type concentration in the emitter. The smoking gun was an enormous difference in the emitter sheet resistance ($R_{she}$) between the single junction solar cell (SJSC) case and the top subcell of a dual-junction solar cell (DJSC) case, from 450 to 1600 Ω/sq, despite the same growth parameters were used. This enormous increment in $R_{she}$ led to a high total series resistance that turned the solar cell unpractical for concentration applications.

Different cases related to the out-diffusion of Zn in III-V semiconductors devices and, particularly, in multijunction solar cells (MJSC) have been extensively reported in the literature. The doping alteration caused by the movement of Zn across the semiconductor structure has proven to spoil the performance of the device in many ways. For instance, by the introduction of majority carrier barriers which increase the total series resistance of the device [3] or by diminishing the effectiveness of the back surface passivation and thus reducing the collection efficiency of the cell [4][5]. It has been observed that Zn-diffusion is enhanced during the growth of heavily doped n-type layers due to the introduction of a high concentration of point defects. In this sense, the widely accepted mechanism proposed by Deppe in a GaAs/AlGaAs system [6] accounts for a Zn-Ga interaction as the driving force for the Zn-diffusion which takes place in the semiconductor structure. Particularly, this model suggests that the growth of heavily doped n-type layers promotes the generation of an excess concentration of interstitial Ga ($I^{+}_{Ga}$) which diffuses out of the n-type layer. Then, $I^{+}_{Ga}$ kicks-out the substitutional Zn ($Zn_{Ga}$), responsible of the electrically active p-type doping, generating interstitial Zn ($I^{+}_{Zn}$) and locally reducing the free-hole concentration. Once in interstitial positions, Zn is known to be a fast diffuser.

In this work we present strong evidence that pinpoints the tunnel junction growth as the trigger for Zn-diffusion in inverted MJSCs. Specifically, the growth of the 25 nm n++ GaAs cathode after a p+ Zn-doped AlGaInP BSF layer leads to a Zn-redistribution from the BSF towards the emitter direction in the GaInP top cell. We discuss this finding through electrochemical capacitance voltage (ECV) measurements in a GaInP/AlGaInP/GaAs system, corresponding to the top subcell and the nearby tunnel junction (TJ) of an IMM. The results obtained allow us to infer some issues related to the thermodynamics of point defects during the growth process in accordance with the doping profiles observed in GaInP solar cells grown in an upright configuration. Here, diffusion is triggered by the growth of the heavily doped front contact layer, as observed in other works [4]. Furthermore, the Zn-diffusion process and the altered pn junction profiles have an impact on the bandgap of the TC, through the reduction of the degree of CuPt ordering; on the rear-passivation of the GaInP/AlGaInP interface and on the bulk minority carrier properties. These implications are discussed by contrasting doping profile evolutions and the corresponding QE and I-V curves.

## II. METHODS

Several structures corresponding to different subsets of an inverted double junction GaInP/GaAs solar cell were grown on GaAs substrates with 2(111)B miscut in an horizontal low-pressure MOVPE reactor (AIX200/4). The precursors used were $AsH_3$, $PH_3$ for group-V, TMGa and TMIn for group-III and DTBSi, DETe, $CBr_4$ and DMZn for dopant elements. All growth conditions, including thicknesses, growth rates, III/V ratios, phosphine partial pressures and molar flows, were kept constant in order to obtain fair comparisons. The nominal doping levels in the as-grown TC structure were confirmed by comparing SIMS and ECV measurements with a calibrated structure. For simplicity, the growth direction of all samples is assumed to be inverted, unless specified otherwise.

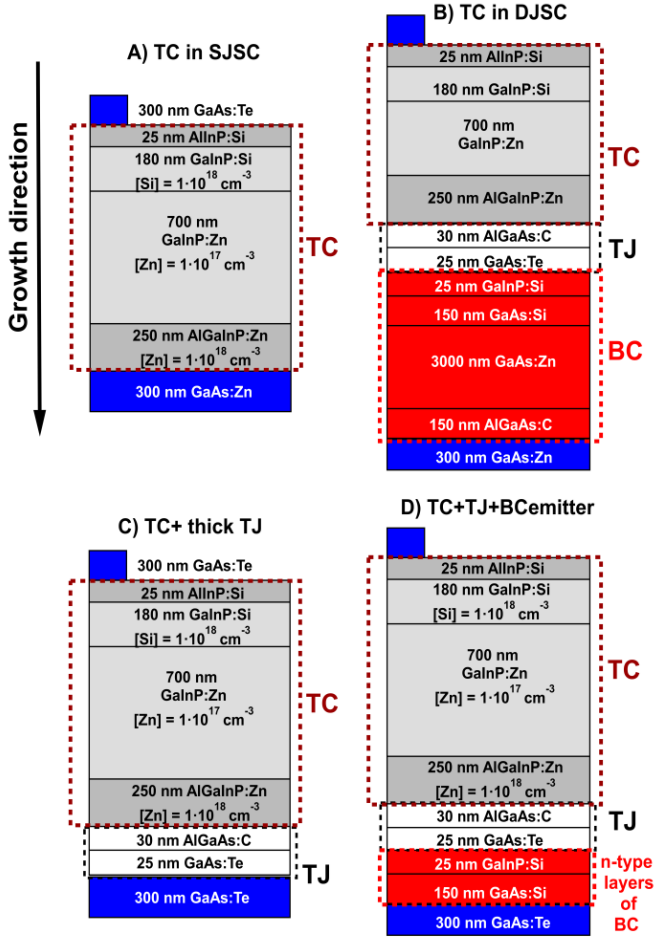

Fig. 1. Sketch of the TCs layers of the different samples under study corresponding to subsets of a 2J GaInP/GaAs inverted solar cell: A) TC in a SJSC; B) TC in a DJSC; C) TC+thick TJ and D) TC+TJ+BC emitter.

The goal of this study is to figure out how Zn-diffusion evolves throughout the different stages of the epitaxial growth as well as analyze the role played by different parameters, like the growth direction or the material used in the cathode of the TJ. For this purpose, GaInP subcells of a single-junction and a dual-junction employing a GaAs/AlGaAs TJ [7], referred to as **TC in SJSC** and **TC DJSC** and sketched in Fig.1, are used as benchmarks in the study. Then, different samples comprising intermediate structures and thermal loads are considered:

- **Case A**: A GaInP SJSC sample which has been subjected to a MOVPE annealing under arsine simulating the growth of the whole DJSC, corresponding to 75 min at 675ºC (**TC+annealing**).
- **Case B**: A GaInP SJSC which incorporates the GaAs/AlGaAs TJ, the window and the emitter of the GaAs bottom subcell (**TC+TJ+BCemitter** in Fig.1D).
- **Case C**: A GaInP/GaAs DJSC with GaInP instead of GaAs in the TJ cathode, referred to as **TC in DJSC (GaInP cathode)** [8]
- **Case D**: A GaInP SJSC incorporating the GaAs/AlGaAs tunnel junction and a contact layer identical to the cathode of the TJ (**TC+thick TJ** in Fig.1C).
- **Case E**: A GaInP SJSC identical to the represented in Fig.1A but grown in the opposite direction, from p-type to n-type layers (**TC in SJSC upright**).

Solar cell devices from the different samples were fabricated using a similar inverted process as described in [9]. Electroplated gold was used for p-type and n-type ohmic contacts and no antireflection coatings were applied. Solar cells were characterized by measuring dark and light I-V curves and quantum efficiency (QE), whereas the emitter sheet resistance was obtained by the Van der Pauw method. A chemical etching up to the AlGaInP BSF layer of the TC was performed in all samples previous to the ECV measurements of the TC layers.

## III. RESULTS AND DISCUSSION

### A. Doping profiles

Some information about the diffusion mechanism is provided by the contrast of the ECV measurements presented in Fig.2, comprising the cases enumerated in the previous section. In this figure, the free carrier concentration profile in the TC of the benchmark structures is included in all panels A to E (as solid lines for the SJSC and as dashed lines for the DJSC).

The first point to make about the benchmark structures is the absence of Zn-diffusion enhancement in the top subcell of the as-grown SJSC (TC in SJSC), i.e. when no n-type layers are grown after the BSF layer. In this case, the nominal doping level in the BSF is attained ($1 \cdot 10^{18}$ cm$^3$) and, more important, the doping profile remains almost unaltered when the structure is submitted to a thermal annealing equivalent to the growth of the DJSC (round symbols in Fig.2A). It was also observed that $R_{she}$ stayed constant at a value of 450 Ω/sq. This behavior unequivocally excludes the thermal load as the unique cause of the Zn-diffusion enhancement. In other words, the conditions which favor Zn-diffusion are not settled yet in this structure and, as a consequence, the thermal energy provided by the MOVPE annealing has no further impact in the Zn distribution.

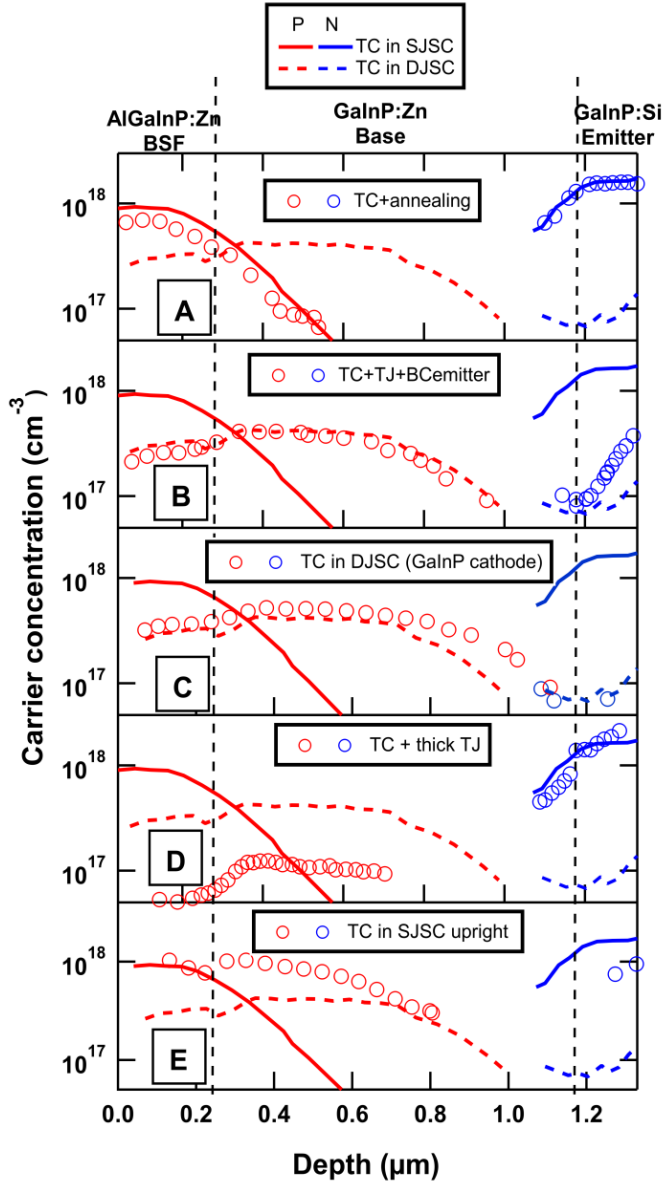

Fig. 2. Comparison of the different doping profiles of the TC layers. An as-grown SJSC and a DJSC are used as benchmark and plotted in all cases, while the third profile is different for A (TC+annealing); B (TC+TJ+BCemitter), C (TC in DJSC with GaInP cathode), D (TC+ thick TJ) and E (TC in SJSC upright).

On the contrary, when the GaAs subcell is grown after the top subcell, giving rise to the dual-junction cell (DJSC in the same graph), a strong Zn-diffusion from the BSF to the base is produced, leading to a drastic drop in the n-type carrier concentration of the emitter. This diffusion is assumed to be the origin of the high $R_{she}$ observed, around 1600 Ω/sq. It is significant how the profile of the DJSC matches quite well to that obtained when only the first nanometers of the GaAs subcell are grown (Fig.2B), with a high quantity of Zn migrating from the BSF to the absorber layer too. In this regard, the thermal load suffered by the BSF layer in the TC+TJ+BCemitter is very low in comparison with the DJSC (17 minutes vs 90 minutes at ≈ 675ºC), suggesting that the TJ growth is indeed catalyzing the effect of the thermal load on the rapid Zn diffusion. Interestingly, as shown in Fig.2C, the diffusion enhancement is similar when the tunnel junction cathode employs GaInP. This suggests that the mechanism of point defects injection and diffusion is analogous for both materials. Experiments to understand these mechanisms are underway.

Fig.2D provides a further insight: with a thick GaAs contact layer (identic to the cathode) grown immediately after the TJ, a strong variation in the Zn-concentration of the BSF occurs (from $1·10^{18}$ cm$^{-3}$ to $5·10^{16}$ cm$^{-3}$), therefore, pointing out the cathode growth as the origin of the enhanced Zn diffusion. The obtained profile suggests that Zn is moving towards the TJ, as the base doping level is kept constant at nominal values of $1·10^{17}$ cm$^{-3}$, and possibly damaging the electrical performance of the TJ [10]. In fact, we have observed a TJ failure in those samples at around 600 suns when are submitted to an additional MOVPE annealing.

Finally, a revealing heavy diffusion is also produced when the single-junction GaInP cell is grown in the upright direction (Fig.2E). In this case, Zn-diffusion would be induced by the growth of the n++ front contact layer, as reported in other works [4]. This fact was confirmed by growing the same structure with an undoped base and obtaining the same profile. Thus, the high Zn-concentration in the base, observed as a high free-hole concentration, must come from the BSF layer.

Considering all these facts together, the link between the growth of n-type layers and Zn-diffusion becomes clear. In addition, as both the inverted and upright SJSC incorporate a heavily doped front contact layer, but only the upright one presents a remarkable Zn-diffusion, it can be inferred that such interaction occurs only when the n-type layers are grown after Zn-doped layers. This fact seems to suggest that point defects generated during the growth of heavily n-type doped layers rapidly diffuse and annihilate by Frenkel reactions reaching their equilibrium concentration. Furthermore, the diffusion observed in the upright solar cell points-out that point defects generated by the growth of the n++ front contact layer can diffuse at least 1 μm until reaching the buried BSF layer, during the short growth time of the contact layer (4 minutes). In this way, the underlying thermodynamic mechanism is highly dependent on the growth direction and should be contemplated when designing an inverted solar cell.

## B. Solar cell performance

The altered pn junctions profiles will modify the performance of solar cells in different ways, depending on how it is produced. In this work, we account for different cases: a bandgap increase in GaInP, a poor passivation at the GaInP/AlGaInP interface and a deterioration of the bulk minority carrier properties.

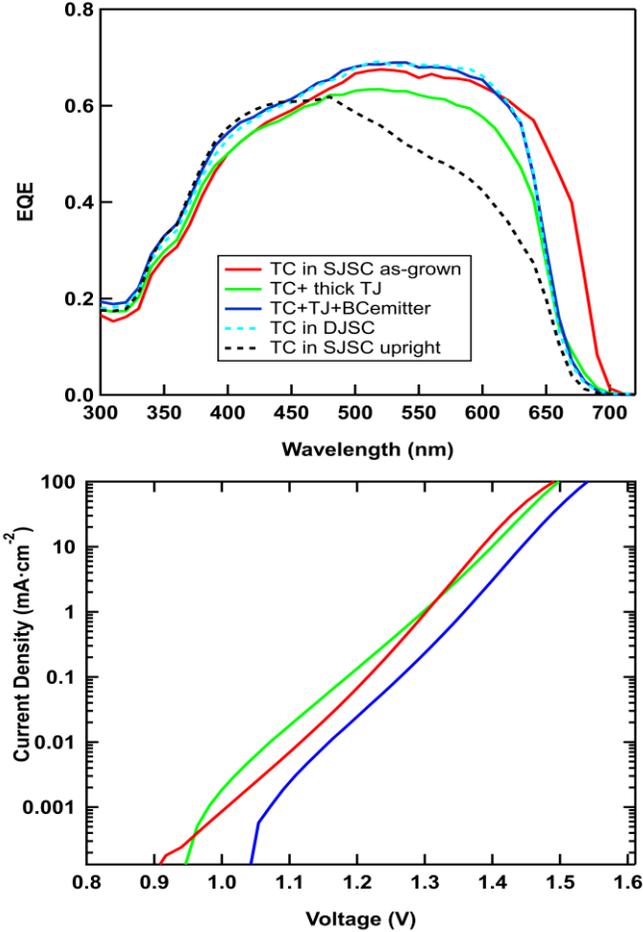

Fig. 3. Quantum efficiency (top) and dark-IV curves (bottom) of different TCs corresponding to different Zn-diffusion cases.

First, it must be considered that Zn impurities become electrical active when are located in the group-III sites of the crystal lattice and thus, substituting group-III elements. For instance, a Zn atom introduces a free hole into GaInP by replacing either a Ga or an In atom in the cation site of the lattice. In this way, doping profile alterations observed from ECV measurements (Fig.2) necessarily imply the presence of interactions in the group-III lattice sites via kick-out processes that rearrange Ga and In atoms positions in the bulk. As a consequence, the degree of ordering in the highly ordered as-grown GaInP [11] is susceptible to be modified by Zn-diffusion across the bulk, producing a consequent bandgap blue-shift. Besides, the higher the concentration of Zn moving around, the stronger the disordering effect should be. In this regard, a bandgap increase can be directly perceived from the lower wavelength cut-off in the EQE strictly in samples where a strong Zn-diffusion takes place (TC+thick TJ; TC+TJ+BCemitter; TC in DJSC and TC in SJSC upright), demonstrating that, indeed, the Zn-diffusion process promotes disordering in the GaInP crystal lattice (Fig. 3).

Additionally, the severe drop in the doping level of the BSF in the TC+thick TJ sample (Fig.2D) leads to an important negative doping offset in the base/BSF interface which diminish the rear passivation effectiveness. Consequently, a lower QE response is obtained, more pronounced at long wavelengths, as seen in Fig. 3. We also observed that collection efficiency in this sample is more dependent on the electrical field than the other solar cells, resulting in significantly improved QEs when reverse voltages bias are applied, as observed in other works too [4].

Furthermore, Zn-diffusion can potentially deteriorate minority carrier diffusion lengths if an excess of Zn accumulates in the absorber layer. This is the case of the upright GaInP SJSC, where the high injection of Zn from the BSF to the base (Fig.2E) results in a high impurity concentration in the base (around $1 \cdot 10^{18}$ cm$^{-3}$ along 400 nm) and a drastic deterioration of its collection efficiency.

Finally, GaInP disordering and the deterioration of the rear-passivation influence the dark I-V curves too. On the one hand, the higher bandgap of the sample TC+TJ+BCemitter results in a lower recombination current than the TC in a SJSC (higher voltages for the same currents in Fig.3, bottom panel). On the other hand, this bandgap shift benefit cannot be appreciated in the TC+thick TJ, as it may be counterbalanced by a less effective rear-passivation and, as a consequence, a higher recombination current.

## IV. SUMMARY

Zn-diffusion taking place during the MOVPE growth of an IMM has been assessed through the comparison of various samples corresponding to different subsets of an inverted double-junction GaInP/GaAs solar cell; an inverted single-junction GaInP solar cell subjected to a MOVPE annealing and an upright single-junction GaInP solar cell. In this way, the analysis of doping profiles has shown that Zn-diffusion is triggered by the growth of n++ layers, but only when are grown after the Zn containing layer, revealing the mediation of injected point defects. Furthermore, the consequent Zn migration potentially alters many solar cell properties like the bandgap, the GaInP/AlGaInP passivation and the minority carrier diffusion lengths. In this regard, doping profiles obtained in different devices have been related to the resulting quantum efficiencies and dark I-V curves. It becomes clear that these diffusion processes need to be considered when designing multijunction solar cells which inevitably involve the use of Zn and heavily n-type doped layers.

ACKNOWLEDGMENTS

This project has been funded by the Spanish MINECO with the project TEC2017-83447-P and by the Comunidad de Madrid with the project with reference MADRID-PV2 (S2018/EMT-4308). M. Hinojosa is funded by the Spanish MECD through a FPU grant (FPU-15/03436) and I. García is funded by the Spanish Programa Estatal de Promoción del Talento y su Empleabilidad through a Ramón y Cajal grant (RYC-2014-15621).